\documentclass[letterpaper, 10 pt, conference]{ieeeconf}  

\IEEEoverridecommandlockouts                              

\title{\LARGE \bf
PREGNANCY LOSS AND UNETHICAL ALGORITHMS: ETHICAL ISSUES IN
TARGETED ADVERTISING
}

\author{Fatemeh Golpayegani$^{1}$  
\thanks{$^{1}$Fatemeh is with School of Computer Science,
        University College Dublin, Dublin, Ireland
        {\tt\small fatemeh.golpayegani@ucd.ie}}%
}

\begin{document}

\maketitle
\thispagestyle{empty}
\pagestyle{empty}




\section{EXTENDED ABSTRACT}

Pregnancy loss which happens for one in four pregnancies can be a traumatic experience that impacts
women’s personal and social life in several ways. Studies have shown that women who have
experienced pregnancy loss are at a great risk of depression and anxiety and they are more likely to
have major social difficulties. This can be followed by social life distortion and isolation leaving social
media and disclosure to anonymous online support groups a way to grief as well as seeking some help.
However, so-called intelligent targeted advertising will follow them on every on-line media. Targeted
advertisement on baby products or pregnancy clothes can target pregnant women as soon as they do
a few searches on the web, or when the first pregnancy tracking app is installed. Although targeted
advertising algorithms are intelligent enough to identify pregnant women in very early stages through
their search history, the algorithms fail to adapt their mechanism when a pregnancy loss occurs.
Despite all the search history on pregnancy loss symptom and reasons, and activities such as joining
support groups with distinctive names, the algorithms continue to target these women . It will be a
false claim that recognising such situation is not possible despite all the advancement in the field of
social media behaviour tracking, emotion analysis, and history of the on-line activities.
Extensive effort and research has been conducted on ethics in advertisement to regulate the content
and reach of advertising to protect the public over the past years. This includes ethical issues in
controversial advertising, matters of deception, representation, targeting of vulnerable population.
However, there is no particular research on how these issues will be addressed when pregnant women
are targeted and how algorithms can be designed to behave differently when they figure out that a
loss has happened.

\section{ETHICAL ISSUES OF TARGETED ADVERTISING ON PREGNANT WOMEN}

It is common amongst pregnant women to use on-line resources, pregnancy-tracking Apps \cite{rodger2013pregnant} and social media to access pregnancy-related information \cite{zhu2019pregnancy, robinson2014women}. Such resources remain as primary resources that women use even after a pregnancy loss or miscarriage \cite{andalibi2018not}. Pregnancy is not a similar experience for all women, which can depend on many factors such as the family support, women economic and social support, or the burden of their jobs. Both physical changes and the whole unknown that a new child
will bring to a woman’s life sometimes makes pregnancy an overwhelming experience. The targeted advertising can make this experience even worse by featuring glowing in shape pregnant women, or pregnant mothers with their kids in clean and organised houses. Such ads will start as soon as the first pregnancy tracking app is installed or after a few searches on the internet, and will follow the pregnant women recommending them Yoga classes, anti-stretchmarks body oils and towards the end of the pregnancy showing off all the beautiful nurseries and baby products. This might seem intelligent as the
advertisement algorithm will keep the product recommendation updated as the pregnancy
progresses, however this will be a torturous experience for a woman who has lost a pregnancy but keeps receiving all of these updated targeted ads. Many of the general ethical issues such as gender stereotypes \cite{plakoyiannaki2008images}, and female role portrayals \cite{tuncay2015advertising,grau2016gender}, of targeted advertising on women apply to pregnant women as well. Although the literature studies many of such ethical issues raised by advertisement
targeted at women, they lack the analysis of such advertisements targeted at pregnant women. This includes both general products and specific maternity or baby products targeted at pregnant women. The very important question to be asked is why these ads keep following women experiencing a loss, despite the fact that it is easy to identify individuals experiencing pregnancy loss or miscarriage \cite{pang2018online},
either through their social media activity or their search history. This question would not be raised in the first place if the algorithms deciding who to target and what ad to show were designed ethically. Algorithms are designed to learn about us and structure our lives in several ways such as targeted advertisement, and determining our search results which might follow a decision that we will make \cite{martin2019ethical}.
Such algorithms are continuously learning about us through the available Big Data. For example, they can know people’s ethnicity, gender and religion, if somebody is getting a loan, has got fired \cite{o2016weapons}, the places that has already visited or going to visit, or even more personal information such as if getting married or pregnant. Although algorithms can shape our lives in many beneficial ways, their ethical implications must be discussed \cite{mittelstadt2016ethics}, especially for more delicate cases such as advertisements targeted
at pregnant women. Advertising algorithms are getting more accurate having access to more and more data and computational resources to process those data \cite{thompson2018commercial, naor2018near}. They aim to target the best advertisement to the best audience to maximize businesses’ profit. However, they raise multiple ethical issues that has to be considered. As the literature suggests there are several types of ethical concerns that can be raised by algorithms: 
\begin{itemize}
    \item Traceability, how a result of an algorithm can be traced to its source.
    \item Transformative effects, how the original data is transformed to derive to a particular result.
    \item Unfair outcomes, an algorithm can produce biased results.
    \item Misguided evidence, as the produced evidence will depend on the quality of the input data.
    \item Inscrutable evidence, as often it is not possible to understand the relation between the data point being used and the drawn conclusion when machine learning algorithms and big data are used.
    \item Inconclusive evidence, as they are probable but uncertain knowledge that are derived from machine learning or inferential statistics.
    
\end{itemize}

Additionally, the following concerns must be particularly addressed in this context:

\begin{itemize}
    \item Ads’ timeliness, how relevant is the outcome of the algorithm during the time, for example a pregnant woman might experience a very early miscarriage, should the ads go on for months without taking into account any further data or user activity?
    \item Ads’ reach, who is seeing the ads? For example, Should the ads featuring active glowing pregnant women be targeted at the ones experiencing complications during their pregnancy? This also concerns about the reach of the inappropriate ads to single mothers, or ones with weaker social or economical situations.
\end{itemize}

\section{CONCLUSION AND FUTURE WORK:} In this paper, the ethical issues and the importance of ethical algorithm design for target ads were briefly discussed. The targeted ads can become helpful tools if they can recognise cases of loss, and  recommending relevant mind or body care for women experiencing a loss. This requires the algorithms to always adapt and learn from the changes and  use the feedback provided by the user either directly or indirectly through search histories or online activities.


\end{document}